\documentclass[rnote]{aa}
\usepackage[varg]{txfonts}
\usepackage{natbib,twoopt}
\bibpunct{(}{)}{;}{a}{}{,}
\pdfoutput=1

\begin{document}
\title{Merged or monolithic? Using machine-learning to reconstruct the dynamical history of simulated star clusters}
\titlerunning{Machine-learning star-cluster dynamical history}
\author{Mario Pasquato \inst{1} \and Chul Chung \inst{1}}
\authorrunning{Mario Pasquato \and Chul Chung}
\institute{Department of Astronomy \& Center for Galaxy Evolution Research, Yonsei University, Seoul 120-749, Republic of Korea}
\date{Received XXXX/ Received YYYY}

\abstract
{{Machine-Learning (ML) solves problems by learning patterns from data, with limited or no human guidance. In Astronomy, it is mainly applied to large observational datasets, e.g. for morphological galaxy classification.}}
{{We apply ML to gravitational N-body simulations of star clusters that are either formed by merging two progenitors or evolved in isolation, planning to later identify Globular Clusters (GCs) that may have a history of merging from observational data.}}
{{We create mock-observations from simulated GCs, from which we measure a set of parameters (also called \emph{features} in the machine-learning field). After dimensionality reduction on the feature space, the resulting datapoints are fed to various classification algorithms. Using repeated random subsampling validation we check whether the groups identified by the algorithms correspond to the underlying physical distinction between mergers and monolithically evolved simulations.}}
{{The three algorithms we considered (C5.0 trees, k-nearest neighbour, and support-vector machines) all achieve a test misclassification rate of about $10\%$ without parameter tuning, with support-vector machines slightly outperforming the others. The first principal component of feature space correlates with cluster concentration. If we exclude it from the regression, the performance of the algorithms is only slightly reduced.}}
{}


 \maketitle

\section{Introduction}
{Machine-Learning (ML) algorithms are used to automatically extract patterns from datasets and to make predictions based on the acquired knowledge.
They are often used when it is impractical or otherwise not advisable to proceed manually, either because of the size of the data or because of its complexity (e.g. extremely high dimensionality), that prevents the use of simple models and direct visualization. ML algorithms are usually divided into two classes: supervised (in which a set of training examples is used by the algorithm to build a predictive model to be used in a subsequent phase) and unsupervised (in which the algorithm identifies patterns directly from the data).}

{ML techniques are by now ubiquitous in Astronomy, where they have been successfully applied to photometric redshift estimation in large surveys such as the Sloan Digital Sky Survey \citep[][]{2003LNCS.2859..226T, 2006astro.ph.12749L, 2007ApJ...663..774B, 2010ApJ...715..823G, 2011PASP..123..615S, 2012MNRAS.419.2633G, 2013MNRAS.432.1483C, 2014arXiv1406.3192C, 2015arXiv150308214H, 2015MNRAS.449.1275H}, automatic identification of QSOs \cite[][]{2009arXiv0910.3770Y}, galaxy morphology classification \citep[][]{2010MNRAS.406..342B, 2013A&C.....2...67S, 2014PASP..126..959K}, detection of HI bubbles in the interstellar medium \citep[][]{1998A&A...332..429T, 2003PASP..115..662D}, classification of diffuse interstellar bands in the Milky Way \citep[][]{2015arXiv150104631B}, prediction of solar flares \citep[][]{2009SpWea...7.6001C, 2009SoPh..255...91Y}, automated classification of astronomical transients and detection of variability \cite[][]{2008AN....329..288M, 2012arXiv1209.1681D, 2013MNRAS.435.1047B, 2014arXiv1407.4118D, 2015MNRAS.449..451W}, cataloging of impact craters on Mars \citep[][]{2009Icar..203...77S}, prediction of galaxy halo occupancy in cosmological simulations \cite[][]{2013ApJ...772..147X}, dynamical mass measurement of galaxy clusters \citep[][]{2015ApJ...803...50N}, and supernova identification in supernova searches \citep[][]{2007ApJ...665.1246B}. Software tools developed specifically for astronomy are also becoming available to the community, still mainly with large observational datasets in mind \citep[][]{2012cidu.conf...47V, 2014AAS...22325301V, 2014ascl.soft07018V, 2014era..conf30402B}.}

{The goal of this note is to test the feasibility of using ML methods for the automatic interpretation of gravitational simulations of star clusters. In particular, we want to check whether classification algorithms applied to a set of mock-observations obtained from gravitational N-body simulations are able to extract useful information on the underlying physics of the simulations. In order to avoid unnecessary abstraction, we chose to address an actual question (with a yes/no answer) regarding the dynamical history of Globular Clusters (GCs): are some GCs the product of a merger of progenitors or did all GCs evolve in isolation from a monolithic protocluster?}

According to the current consensus, galaxy mergers are quite frequent on the cosmological timescale and probably the main engine of galaxy evolution {\citep[e.g. see][]{1972ApJ...178..623T, 1977egsp.conf..401T, 1994ApJ...437L..47M}. In principle, galaxy merging could also be studied by applying machine-learning to simulations. However the case of GCs is more familiar to the authors and mock-observations are easier to build because most GCs are resolved into stars, so for our discussion we do not need to convert the positions and masses of the simulated stars into a luminosity density profile. From the point of view of running simulations, the fact that GCs lack gas (though they may have contained considerable quantities of it in the past), partly justifies our choice of modelling only dissipationless dynamics with pure particle N-body simulations. On the other hand, in order to simulate a wet galaxy merger the dissipative evolution of the gas component needs to be modeled with hydrodynamical codes, and merger-induced star formation must be taken into account. It has already been suggested that some GCs} may have an history of merging, possibly resulting in massive objects with a strong metallicity spread such as Omega Centauri \citep[][]{1989PASJ...41.1117S, 1996ApJ...471L..31V, 1997ApJ...478L..99C, 2002ASPC..265..337T, 2013MNRAS.435..809A, 2013ApJ...778L..13L} or in nuclear star clusters \citep[][]{2005HiA....13R.381C, 2006ApJ...644..940M, 2008ApJ...681.1136C, 2013MmSAI..84..167C}. Observationally, mergers and monolithic clusters ought to be different either in their sky-projected density distribution or in their stellar kinematics, or both. Dynamical relaxation is bound to erase the initial conditions and the memory of a merger with them, but not necessarily quickly and completely, since it works on relaxation timescales of some Gyr for most GCs \citep[][]{1996AJ....112.1487H, 2005ApJS..161..304M}. Besides an elongated shape due to residual pressure-tensor anisotropy left over by the merger or lingering rotation in the case of off-axis mergers, the clues are probably subtle and difficult to parameterize. {Moreover, deviations from spherical symmetry in clusters are actually observed \citep[][]{1987ApJ...317..246W, 2010ApJ...721.1790C} but there is currently limited consensus as to their explanation, because rotation and tidal effects may result in elongated profiles without the need for a merger \citep[][]{1986A&A...166..177D, 2008ApJ...689.1005B, 2009ApJ...703.1911V, 2012A&A...540A..94V, 2013ApJ...772...67B, 2014MNRAS.443L..79V}.}

In Sect.~\ref{mlwhyhow} we discuss why the merging problem and similar problems related to the interpretation of numerical simulations of star-cluster dynamics are amenable to be tackled by machine-learning, and {how we proceed to use machine-learning to interpret simulations}. In Sect.~\ref{simu} we describe our sets of simulations and provide further details on how the simulations are turned into mock-observations and later classified by the chosen algorithms. In Sect.~\ref{reco} we present our results and conclusions.

\section{Machine-learning: the why and the how}
\label{mlwhyhow}
{The approach we are testing in this note is motivated by looking at the usual accepted procedures employed for running numerical experiments. Namely:
\begin{enumerate}
\item identify a specific astronomical question that can in principle be answered by observations, (H)
\item run simulations for two alternative scenarios, in which the question is answered in the positive or in the negative, respectively, (A)
\item identify a parameter that significantly differs between the two scenarios by examining mock-observations obtained from the simulations, (H\textbf{/A?})
\item (if possible) measure the parameter on actual observations and draw conclusions as to which scenario is more likely. (H/A)
\end{enumerate}
In parentheses we marked with an A the steps that are fully automated, and with an H the steps that require human intervention. The point of this note is to show that the third step in the procedure above can be automated by machine-learning instead of being carried out by manually finding a parameter that discriminates between the two scenarios. The advantage is that subjective, error-prone, and time-consuming human understanding is substituted by an automated procedure. This is an important advantage in the case of complex systems such as star-clusters, where an intuitive grasp of the underlying physics is difficult at best. It may be useful to compare this problem to face recognition, another field where machine-learning algorithms are quite successful, but humans are successful too: no explicit subject-matter knowledge (i.e. detailed knowledge of the human anatomy) is needed for the task of face recognition, neither by humans nor by machine-learning algorithms. On the other hand, humans cannot answer questions regarding the dynamics of star clusters at a glance, without resorting to detailed dynamical models that need to be developed on a case-by-case basis. This is a strong argument to apply machine-learning to this sort of dynamical problems: to automate the interpretation of simulations, which until now is always exclusively manual, despite the simulations themselves being fully automated.
}

\subsection{Reducing our scientific question to a supervised classification problem}
\begin{figure}
\includegraphics[width=0.99\columnwidth]{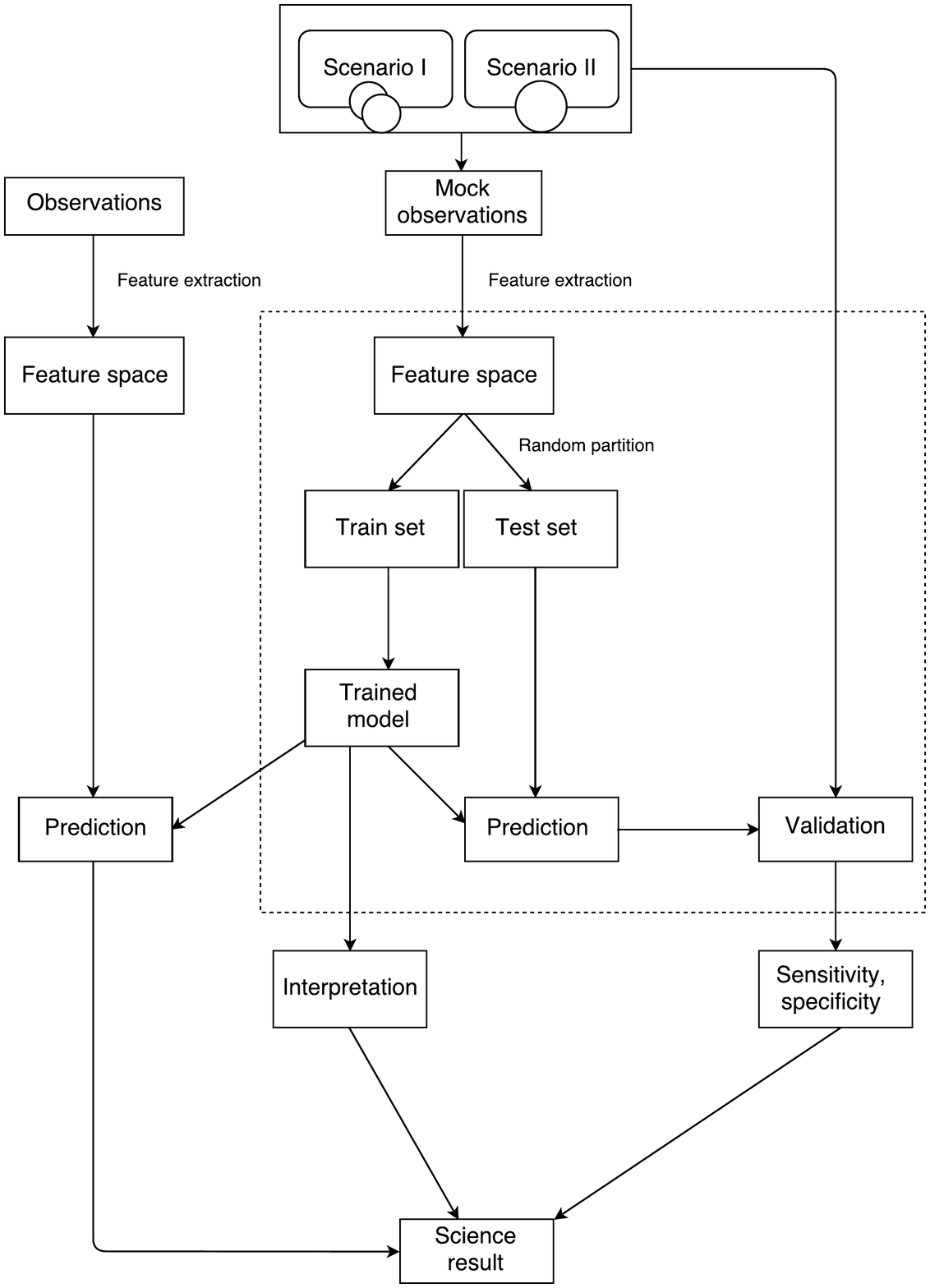}
\caption{Applying supervised classification to the interpretation of simulations, with reference to the GC-merging question. The procedure starts by running simulations of two alternative scenarios, in this case merged GCs or GC evolved in isolation (top center box). Later, mock observations are generated from snapshots and quantitative parameters are extracted, generating the so-called feature space (middle box, following the arrows).  A datapoint in feature space corresponds to a mock observation. The feature space is randomly partitioned in two subsets (train set and test set) containing datapoints corresponding to both scenarios. A model is trained (i.e. has its parameters optimised) on the train set and used to make predictions by classifying the test set, which are later compared with the known ground truth. This is possible because the datapoints in the test set correspond to known simulated scenarios. The process enclosed in the dashed box is repeated several times with different random partitions of the feature space. This results in measures of specificity and sensitivity (true-negative and true-positive rates respectively; bottom right box). The trained model can also be applied to classify actual observations (left column of boxes) and the expected accuracy is known from the validation phase. These ingredients, together with some degree of human interpretation (if possible and/or necessary) may be used to answer the underlying science question (bottom box), provided that the simulations really capture the relevant processes that lead to the observables. \label{supervised}}
\end{figure}

{To pick up signs of past merging in simulations without an ad-hoc modeling of the underlying physical process, we restated the issue as a supervised classification problem. This is applicable in general to any scientific question that requires to discriminate between two (or more) scenarios that can be simulated, producing mock-observations. Figure \ref{supervised} presents the procedure we applied in a visual way.}

{The merged GCs VS isolated GC scenarios can both be easily simulated, as discussed later. The resulting snapshots are turned into mock observations (essentially the sky-projected positions of stars), losing the velocity and line-of-sight position information. This is similar to actual observations of GCs, except for incompleteness and limited field-of-view effects. From mock-observations we extract \emph{features}, i.e. quantitative parameters as described in Sect.~\ref{feat}. The resulting N-dimensional space, where each mock observation generated from a given snapshot corresponds to a datapoint, is called feature space. The classification algorithms attempt to optimise the parameters of a model (`train a model'), essentially drawing a boundary in this multidimensional space. The training step is based on the true classification labels we provide, and the trained model can be later used to classify new datapoints using the learned boundary. There is a wealth of different algorithms to this end, each with different strengths and weaknesses. The algorithms we use are briefly discussed in Sect.~\ref{sup}, but see \cite{hastie01statisticallearning} for a more in-depth discussion. Any algorithm is likely to commit misclassification errors, at least at the theoretical bound set by the Bayes error rate. Hence, a validation step is required to measure accuracy. To this end, we chose to randomly partition the feature space in two subsets (called train set and test set), both containing datapoints corresponding to the two scenarios. A model is trained (i.e. has its parameters optimised) on the train set and used to make predictions on the test set, which are later compared to the known true labels of the test-set datapoints. The process (called validation) is repeated several times with different random partitions of the feature space. This allows us to measure the likelihood of misclassification of each model on this particular dataset, allowing us to obtain an estimate of the accuracy of classification in view of application to actual observations.}

\section{Simulations, mock observations, dimensionality reduction, and learning}
\label{simu}
As we stated above, the point of this note is to provide a proof-of-concept application of machine learning methods to simulated stellar systems, so the complexity of the simulations was tuned down to better focus on the learning aspects of the problem. {Therefore we chose to simulate exclusively dry merging of two equal-mass progenitors\footnote{This is a relatively strong assumption: there is no reason to believe that GC progenitors have to be as gas-free as the GCs we see today, so substantial dissipation and star formation may have taken place in an hypothetic merger event. A dry merger is however chosen for simplicity.}.}

We run {$13$} direct N-body simulations with the state-of-the art direct N-body code NBODY6 \citep[][]{1999PASP..111.1333A}. Initially, they all contain $64000$ equal-mass {single stars, and no primordial binaries except for $3$ simulations that contain only $32000$ stars. Five simulations correspond to the merging of two equal-mass \cite{1966AJ.....71...64K} models of $32000$ stars each, and eight to the monolithic evolution of an isolated King model. All the simulations are evolved for at least 2000} N-body units of time \citep[][]{1986LNP...267..233H}, {which corresponds to about two half-mass relaxation times for the $64000$-star monolitic models. In physical units, if we take the relaxation time of a multi-metallic GC such as $\omega$-Cen to be $\approx 10$ Gyr, this corresponds to about $20$ Gyr, comfortably covering a Hubble time. We do not include tidal interactions with the parent galaxy nor stellar evolution}. The central dimensionless potential of the King models is different across the simulations, in order to explore the effects of different concentrations on the dynamics. Initial conditions are summarized in Tab.~\ref{inico}.

\begin{table}
	\caption{Summary of the initial conditions of our simulations.\label{inico}}
	\centering
	\begin{tabular}{c c c c}
		\hline
		\hline
	 	Simulation & Type & $N$ & $W_0$ \\
		\hline
		32kmerge2$+$2 & Merger & $32000 + 32000$ & $2 + 2$\\
		32kmerge4$+$4 & Merger & $32000 + 32000$ & $4 + 4$\\
		32kmerge7$+$7 & Merger & $32000 + 32000$ & $7 + 7$\\
		32kmerge5$+$2 & Merger & $32000 + 32000$ & $5 + 2$\\
		32kmerge8$+$4 & Merger & $32000 + 32000$ & $8 + 4$\\
		64kW02 & Monol. & $64000$ & $2$\\
		64kW04 & Monol. & $64000$ & $4$\\
		64kW06 & Monol. & $64000$ & $6$\\
		32kW06.5 & Monol. & $32000$ & $6.5$\\		
		64kW07 & Monol. & $64000$ & $7$\\
		32kW07.5 & Monol. & $32000$ & $7.5$\\		
		32kW07.7 & Monol. & $32000$ & $7.7$\\		
		64kW08 & Monol. & $64000$ & $8$\\
		\hline
	\end{tabular}
\end{table}

In the merging simulations, the phase space of each of the two clusters is populated using a different random seed, so that the clusters are not exactly identical. {The clusters are initially set apart by about $30$ times their radius and allowed to fall onto each other with zero initial relative velocity. The reasoning is that leftover rotation from an off-axis merger would make it easier to distinguish a merger from a non-merger, so we want to present our classifiers with a worst-case scenario, meanwhile also limiting the parameter space that we need to probe\footnote{Similarly, increasing the initial relative velocity increases the center-of-mass energy of the collision, enhancing mass-loss (more stars become unbound and leave the cluster right after the merger) and consequently is expected to make the effects of the merger more evident.}. Clearly, in this setup the system's angular momentum is initially $0$,} but some angular momentum can be acquired nonetheless later on, because the system may expel mass slightly asymmetrically. {Head-on merging is nonetheless a simplifying assumption that we plan to relax in a future paper}. {The mergers take place essentially istantaneously (actually on a multiple of the crossing time) while the internal evolution takes place on the relaxation time scale, which is orders of magnitude longer. Therefore the pre-merger internal evolution of the progenitors does not affect the outcomes of the merger simulations.} We extract $100$ subsequent snapshots separated by one N-body unit of time for each simulation, starting at {$1100$ N-body units of time for the $64000$-star models, corresponding to the relaxation time of the 64kW07 model, and at $600$ N-body units of time for the three monolithic, $32000$-star models, corresponding to the relaxation time of the 32kW06.5 model}. 

\subsection{Mock observations and feature extraction}
\label{feat}
\begin{figure}
\includegraphics[width=0.9\columnwidth]{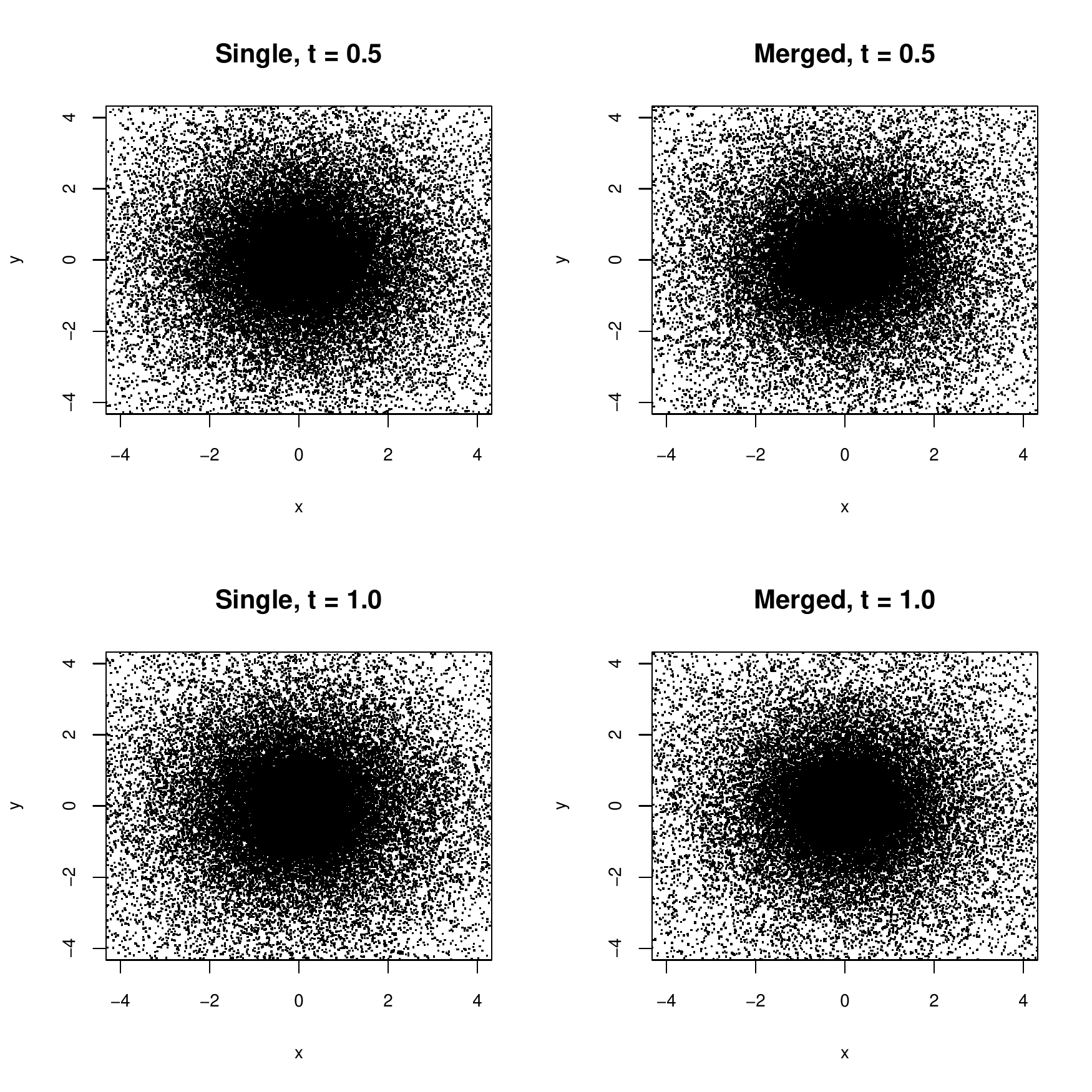}
\caption{Comparison of snapshots taken from a $64000$-star single cluster (64kW08, left column) and a merger of two $32000$-star clusters (32kmerge7$+$7, right column). The rows correspond to different times in units of the single-star cluster half-mass relaxation time. In the plots each point represents a star in the plane of the sky. The x- and y-axis are in units of the projected half-mass radius. It is hard to visually tell the difference between the two snapshots. The bottom panel corresponds to the age-range of the snapshots used for the machine-learning study. \label{human_05_1}}
\end{figure}

\begin{figure}
\includegraphics[width=0.9\columnwidth]{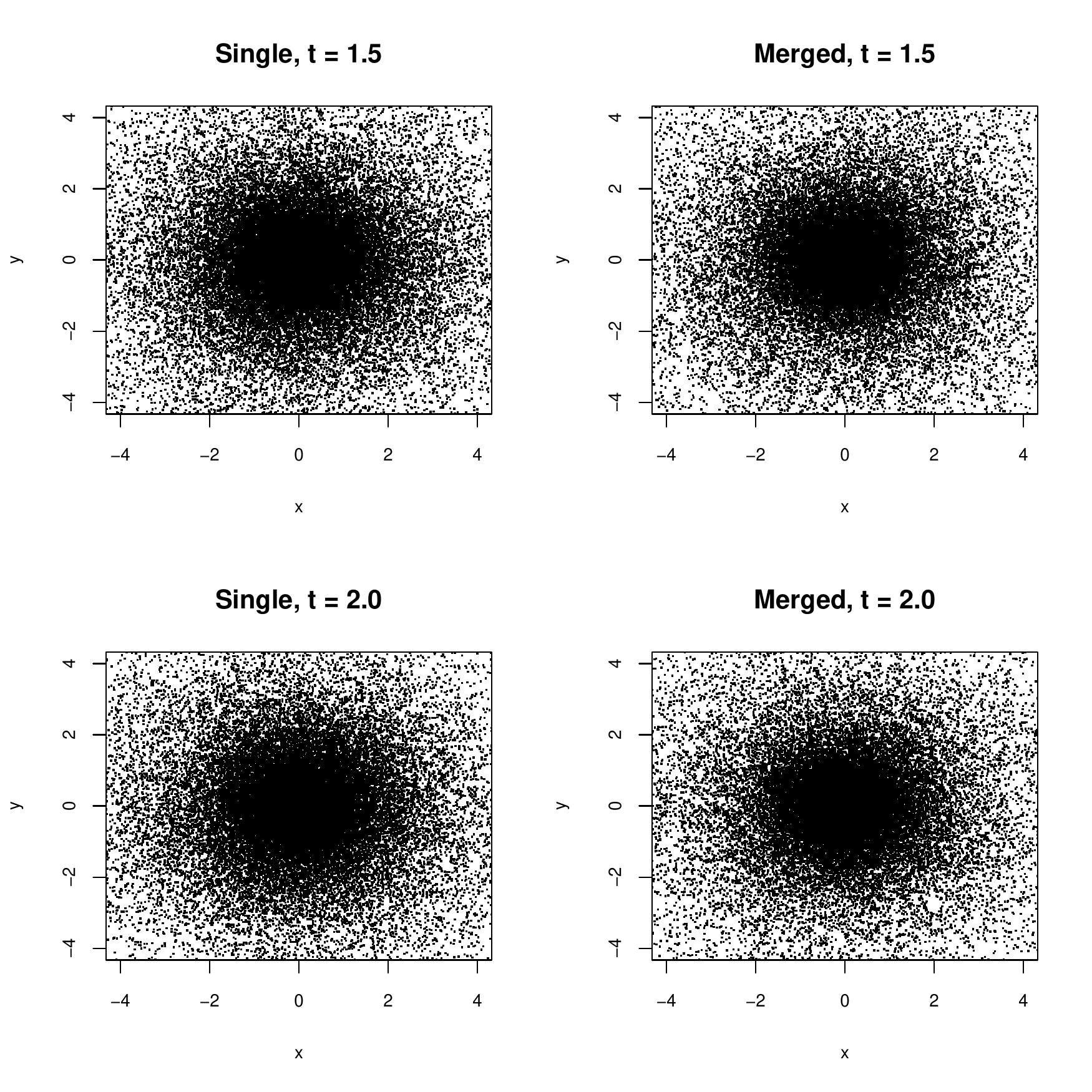}
\caption{Same as Fig.~\ref{human_05_1} for later times.\label{human_15_2}}
\end{figure}

{The snapshots extracted from each simulation are randomly rotated before 2D projection, because we expect that, in actual observations, the plane of the sky should not be in any way privileged in relation to the axis of the merging.} {Two-dimensional sky-projected positions of stars in each snapshot are obtained, ignoring radial velocities, proper motions and the third spatial dimension.} {A finite field-of-view was imposed by discarding stars that fell outside a square of side four times the snapshot's 2D half-mass radius, and observational incompleteness was simulated by randomly extracting only $75\%$ of stars from the snapshots}. {In this context} it would be tempting to use directly as features the $x$ and $y$ sky-projected positions of each star, i.e. to represent each snapshot as $2N$ numbers where $N$ is the number of stars. However this representation would be dependent on the way we sort stars, i.e. a snapshot would be represented differently if its stars underwent a permutation. Since our stars are indistinguishable (having the same mass) this is clearly not desirable. Consequently we decided to use instead suitable quantiles of the $x$ and $y$ variables over the sample of stars of each snapshot. In particular, we chose to represent each snapshot with the deciles ($10\%$ to $90\%$ quantiles) of $x$ and $y$. We centered the deciles by subtracting the median, and standardised them, dividing by the interquartile range, i.e. a measure of dispersion corresponding to the difference between the top and the bottom quartiles. Due to this procedure the median becomes identically $0$, so we do not include it in the final feature space. The centering and standardisation remove from feature space any reference to the absolute position and size of the snapshot\footnote{The standardization is carried out independently in the $x$ and $y$ directions, i.e. both the $x$ and the $y$ interquartile ranges are set to $1$. Thus some additional information about the shape of the cluster is lost.}. {We do this to simulate a realistic comparison between simulations and observations, where the absolute size of a cluster (e.g. in terms of parsecs) depends on initial conditions and cannot be predicted by simulations, which are scale-free. This procedure} leaves us with a $16$-dimensional feature space (eight standardized deciles for each of the two coordinates, which we will denote with $x_{10\%}$ ... $x_{90\%}$ and $y_{10\%}$ ... $y_{90\%}$). Building the feature space in this way also has the advantage that, once the quantiles are calculated, the total number of stars in the snapshot does not enter the calculations any more, so that it is not more computationally expensive to analyze snapshots of simulations with a large number of stars. {The visual appearance of the mock observations is displayed in Fig.~\ref{human_05_1} and Fig.~\ref{human_15_2}, at different evolutionary stages, for the 64kW08 and the 32kmerge7$+$7 models. These models have similar effective concentrations (with reference to the first principal component of the feature space; see discussion in Sect.~\ref{predimred}) and thus were chosen for illustrating the difficulties of telling, by mere eye comparison, if a cluster is merged or monolithic.}

\subsection{Preliminary dimensionality reduction}
\label{predimred}
\begin{figure}
\includegraphics[width=0.99\columnwidth]{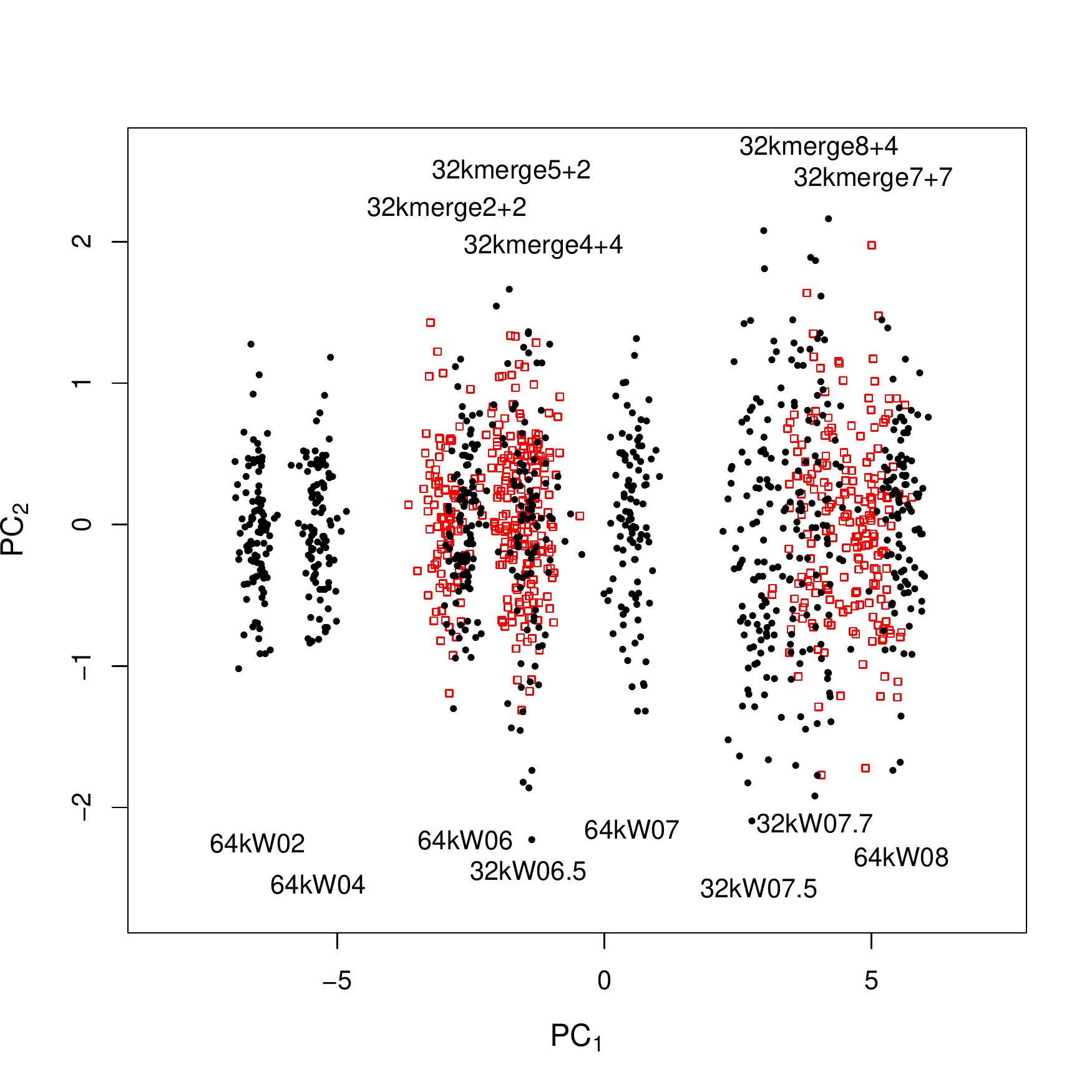}
\caption{Feature space projected on the first two principal components ($PC_1$ on the $x$- and $PC_2$ on the $y$-axis). Each symbol represents a mock-observation obtained from a snapshot. Groups of symbols correspond to simulation runs and are labeled based on the name of each run. Empty red symbols correspond to merger simulations, filled black symbols to monolithic simulations. It appears that $PC_1$ correlates with the initial dimensionless potential $W_0$ of each run. \label{pc12}}
\end{figure}

\begin{figure}
\includegraphics[width=0.99\columnwidth]{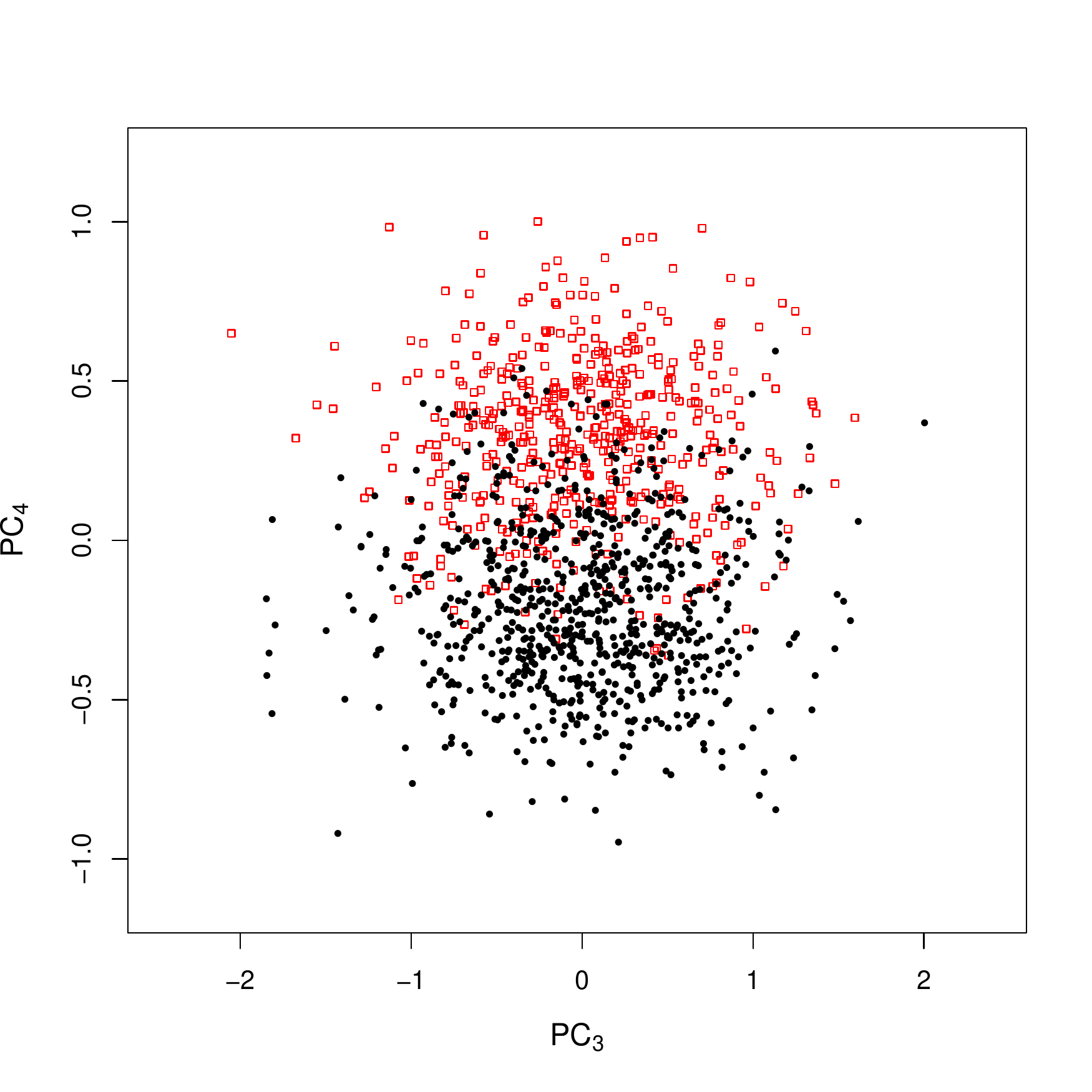}
\caption{Feature space projected on the third and fourth principal components ($PC_3$ on the $x$- and $PC_4$ on the $y$-axis). Like in Fig.~\ref{pc12}, each symbol represents a mock-observation obtained from a snapshot. Different simulation runs are harder to distinguish in this projection, so the labels used in Fig.~\ref{pc12} have been dropped. Empty red symbols correspond to merger simulations, filled black symbols to monolithic simulations. Mergers and monolithic observations appear systematically different in $PC_4$. \label{pc34}}
\end{figure}

{We did not apply machine-learning algorithms directly to the train subset of the feature space. We preliminarily used principal component analysis to obtain the (centered and scaled) principal components of the train set, over which we also projected the test set. This is useful for visualization and interpretation purposes, and it may result in improved accuracy with respect to the original, untransformed coordinates if some of them are irrelevant to the classification problem. We limited the number of components used in the following analysis by omitting components if their standard deviations were less than or equal to $0.05$ times the standard deviation of the first component, by using the R \emph{prcomp} command with a \emph{tol} setting of $0.05$. This results in only the first eight components being retained. In Fig.~\ref{pc12} we plot the first two principal components of the whole feature space, and in Fig~\ref{pc34} the third and fourth. The principal components obtained from the whole feature space differ slightly from the principal components calculated only on the train set, but in the validation phase we used the latter in order to avoid compromising the independence of the test and train data. Still the principal components of the whole feature space can be used for the current qualitative discussion and visualization.}

{We see from Fig.~\ref{pc12} that for monolithic simulations the first principal component $PC_1$ is essentially determined by the cluster's initial $W_0$, that is its concentration. Monolithic simulations with a given initial value of $W_0$ produce snapshots that clump together in $PC_1$, and snapshots from merger simulations also clump around values of $PC_1$ corresponding to effective values of $W_0$. To avoid this clumping effect, a larger number of simulations should be included in \textbf{future studies}, filling in the gaps in $W_0$. Qualitatively, from Fig.~\ref{pc34}, the fourth principal component seems to contribute the most to distinguishing merged from non-merged clusters.}

\subsection{Supervised learning} 
\label{sup}
{We used three different classification algorithms on the feature space in order to compare their performance and accuracy, all in \textbf{the} R language \citep{R} implementation:
\begin{itemize}
\item the $C5.0$ classification-tree algorithm \citep[][R package \emph{C50}]{Quinlan86inductionof, Quinlan:1993:CPM:152181},
\item the k-nearest neighbor algorithm \citep[\textbf{KNN;} see e.g.][R package \emph{class}]{doi:10.1080/00031305.1992.10475879}, and
\item the support-vector machines algorithm \citep[\textbf{SVM;}][R package \emph{e1071}]{svm}.
\end{itemize}
}
{While it is outside of the scope of this note to explain the functioning of the above classification algorithms in detail, we briefly recall their basic functioning in the following.}
{The $C5.0$ algorithm is a particular case of decision tree learning. The purpose of the algorithm is to infer a decision tree based on the training data. The decision tree, whose interior nodes correspond to decisions based on the value of one of the features, will then be used to classify test data. The algorithm works by recursively partitioning the feature space across lines parallel to the coordinates of the feature space. This results in subsets of the feature space that are associated to an end node (leaf) of the tree, yielding a definite prediction for the dependent variable, in this case the classification label. Tree-learning algorithms differ mainly based on how they decide to pick the features on which to split, and on when to stop growing the tree.}
{The k-nearest neighbor algorithm classifies a point in feature space based on the classification of its $k$ nearest neighbors. The Euclidean distance in the transformed feature space (i.e. in the feature space projected on the first two principal components, calculated using the train set only) is used to find the $k$ points of the train set that are nearest to the point for which a prediction is being sought. Then, a majority vote among the $k$ points is used to decide how to classify the target. In this procedure the only adjustable parameter is $k$, and we chose $k = 5$. In a more complex setting than ours, where a larger number of simulations is available and additional variability in the datapoints is introduced by mimicking observational uncertainties more realistically when making mock observations, it may be worthwhile to optimize the value of $k$ through cross-validation to deliver the maximum accuracy of classification. However, this would be premature at this stage.}
{The support-vector machines algorithm is based, in its linear version, on finding the hyperplane that best separates two groups of points in feature space that belong to different classes. Such a hyperplane is, for the purposes of the algorithm, defined as the one for which the distance between the two groups along its perpendicular is maximal. Since linear separation is not always possible, in the general case the points in feature space are mapped to points in a higher-dimensional space where it is easier to find a separating hyperplane, which corresponds to a curved hypersurface when brought back to the original feature space.}

\section{Results}
\label{reco}
{The performance of the algorithms described in \ref{sup} was evaluated by measuring the rate of misclassification defined as follows:}
\begin{equation}
\label{coso}
m = \frac{FP + FN}{N}
\end{equation}
{where $FP$ is the number of false positives, $FN$ the number of false negatives, and $N$ the total number of snapshots.}
{The calculation was repeated over $100$ random splits of the data into equal sized train- and test-sets, and is presented in Tab.~\ref{resu}}

\begin{table}
	\centering
	\caption{Summary of the misclassification rates for the three different algorithms we used (column $1$). The mean value of $m$ as defined in Eq.~\ref{coso} is reported in column $2$, and its standard deviation in column $3$.\label{resu}}
	\begin{tabular}{c c c}
		\hline
		\hline
	 	Algorithm & Mean misclass. rate & Standard deviation\\
		\hline
		C$5.0$ & $0.12$ & $0.02$\\
		KNN ($k=5$) & $0.13$ & $0.01$\\
		SVM & $0.09$ & $0.02$ \\
		\hline
	\end{tabular}
\end{table}

{C$5.0$ (with default settings) and KNN (with $k=5$) perform similarly with slightly more than $10\%$ of the snapshots being misclassified. SVM (also with default settings) has a somewhat higher performance, misclassifying slightly less than $10\%$ of the snapshots.}

{However, we have seen in Sect.~\ref{predimred} that the first principal component $PC_1$ of feature space is essentially representing the initial $W_0$. The initial values of $W_0$ for the monolithic simulations included in this study were arbitrarily chosen to span the range $W_0 = 2 - 8$ resulting in clumps along $PC_1$ which reflect such an arbitrary choice. In the light of this issue, it is interesting to see how the algorithms perform if they are applied to the $({PC}_2, ..., {PC}_8)$ space, excluding $PC_1$. The results are listed in Tab.~\ref{resunopc1}, and show a slight decline in performance.}

\begin{table}
	\centering
	\caption{Summary of the misclassification rates for the three different algorithms (column $1$) after excluding the first principal component from the transformed feature space. The mean value of $m$ as defined in Eq.~\ref{coso} is reported in column $2$, and its standard deviation in column $3$.\label{resunopc1}}
	\begin{tabular}{c c c}
		\hline
		\hline
	 	Algorithm & Mean misclass. rate & Standard deviation\\
		\hline
		C$5.0$ & $0.16$ & $0.02$\\
		KNN ($k=5$) & $0.16$ & $0.02$\\
		SVM & $0.14$ & $0.02$ \\
		\hline
	\end{tabular}
\end{table}

\section{Conclusions}
{There is a great variety of machine learning techniques that can be applied to dynamical questions regarding the evolution of GCs, once they have been reduced to a classification problem. Our ability to obtain scientific results on real data using this method is constrained by our ability to:}
\begin{itemize}
\item {translate complex scientific questions into classification problems}
\item {run simulations of different scenarios that capture all the physical mechanisms that significantly affect the observed data, so that simulated and observed data have the same statistical properties}
\item {build mock-observations that strictly follow the real limitations of actual observational procedures}
\item {solve the actual machine-learning problem with proper application of the algorithms as quantification of the classifier performance}
\end{itemize}
{In this note we demonstrated how the use of simple machine learning algorithms can result in the automatic reconstruction of the dynamical history of simulated systems, at least regarding the specific question we decided to investigate, i.e. is it possible to tell clusters which underwent a merger event from monolitic ones? In our setup, three common machine-learning algorithms managed to achieve a misclassification rate of about $10\%$ without any particular effort in parameter tuning. This is promising for the future application of this method to a larger set of more realistic simulations and eventually observational data. To this end we need to relax our limiting assumptions in the simulation set-up (head-on free-fall merger only, no tidal effects, no primordial binaries, no mass-spectrum) and improve the process that generates mock observations (including incompleteness that depends on distance from the cluster center, a realistic field-of-view, et cetera).}

\section*{Acknowledgements}
This note is a result of the collaborative project between Korea Astronomy and Space Science Institute and Yonsei University through DRC program of Korea Research Council of Fundamental Science and Technology(DRC-12-2-KASI) and from NRF of Korea to CGER. {M.P. acknowledges support from Mid-career Researcher Program (No. 2015-008049) through the National Research Foundation (NRF) of Korea. C.C. acknowledges support from the Research Fellow Program (NRF-2013R1A1A2006053) of the National Research Foundation of Korea.}

\bibliography{manuscript}

\begin{thebibliography}{64}
\expandafter\ifx\csname natexlab\endcsname\relax\def\natexlab#1{#1}\fi

\bibitem[{{Aarseth}(1999)}]{1999PASP..111.1333A}
{Aarseth}, S.~J. 1999, \pasp, 111, 1333

\bibitem[{Altman(1992)}]{doi:10.1080/00031305.1992.10475879}
Altman, N.~S. 1992, The American Statistician, 46, 175

\bibitem[{{Amaro-Seoane} {et~al.}(2013){Amaro-Seoane}, {Konstantinidis},
  {Brem}, \& {Catelan}}]{2013MNRAS.435..809A}
{Amaro-Seoane}, P., {Konstantinidis}, S., {Brem}, P., \& {Catelan}, M. 2013,
  \mnras, 435, 809

\bibitem[{{Bailey} {et~al.}(2007){Bailey}, {Aragon}, {Romano}, {Thomas},
  {Weaver}, \& {Wong}}]{2007ApJ...665.1246B}
{Bailey}, S., {Aragon}, C., {Romano}, R., {et~al.} 2007, \apj, 665, 1246

\bibitem[{{Ball} {et~al.}(2007){Ball}, {Brunner}, {Myers}, {Strand}, {Alberts},
  {Tcheng}, \& {Llor{\`a}}}]{2007ApJ...663..774B}
{Ball}, N.~M., {Brunner}, R.~J., {Myers}, A.~D., {et~al.} 2007, \apj, 663, 774

\bibitem[{{Ball} \& {Gray}(2014)}]{2014era..conf30402B}
{Ball}, N.~M. \& {Gray}, A. 2014, in Exascale Radio Astronomy, 30402

\bibitem[{{Banerji} {et~al.}(2010){Banerji}, {Lahav}, {Lintott}, {Abdalla},
  {Schawinski}, {Bamford}, {Andreescu}, {Murray}, {Raddick}, {Slosar},
  {Szalay}, {Thomas}, \& {Vandenberg}}]{2010MNRAS.406..342B}
{Banerji}, M., {Lahav}, O., {Lintott}, C.~J., {et~al.} 2010, \mnras, 406, 342

\bibitem[{{Baron} {et~al.}(2015){Baron}, {Poznanski}, {Watson}, {Yao}, {Cox},
  \& {Prochaska}}]{2015arXiv150104631B}
{Baron}, D., {Poznanski}, D., {Watson}, D., {et~al.} 2015, ArXiv e-prints

\bibitem[{{Bertin} \& {Varri}(2008)}]{2008ApJ...689.1005B}
{Bertin}, G. \& {Varri}, A.~L. 2008, \apj, 689, 1005

\bibitem[{{Bianchini} {et~al.}(2013){Bianchini}, {Varri}, {Bertin}, \&
  {Zocchi}}]{2013ApJ...772...67B}
{Bianchini}, P., {Varri}, A.~L., {Bertin}, G., \& {Zocchi}, A. 2013, \apj, 772,
  67

\bibitem[{{Brink} {et~al.}(2013){Brink}, {Richards}, {Poznanski}, {Bloom},
  {Rice}, {Negahban}, \& {Wainwright}}]{2013MNRAS.435.1047B}
{Brink}, H., {Richards}, J.~W., {Poznanski}, D., {et~al.} 2013, \mnras, 435,
  1047

\bibitem[{{Capuzzo-Dolcetta}(2013)}]{2013MmSAI..84..167C}
{Capuzzo-Dolcetta}, R. 2013, \memsai, 84, 167

\bibitem[{{Capuzzo-Dolcetta} \& {Miocchi}(2008)}]{2008ApJ...681.1136C}
{Capuzzo-Dolcetta}, R. \& {Miocchi}, P. 2008, \apj, 681, 1136

\bibitem[{{Capuzzo-Dolcetta} {et~al.}(2005){Capuzzo-Dolcetta}, {Di Matteo}, \&
  {Paolini}}]{2005HiA....13R.381C}
{Capuzzo-Dolcetta}, R.~A., {Di Matteo}, P., \& {Paolini}, S. 2005, Highlights
  of Astronomy, 13, 381

\bibitem[{{Carrasco Kind} \& {Brunner}(2013)}]{2013MNRAS.432.1483C}
{Carrasco Kind}, M. \& {Brunner}, R.~J. 2013, \mnras, 432, 1483

\bibitem[{{Catelan}(1997)}]{1997ApJ...478L..99C}
{Catelan}, M. 1997, \apjl, 478, L99

\bibitem[{{Cavuoti} {et~al.}(2014){Cavuoti}, {Brescia}, \&
  {Longo}}]{2014arXiv1406.3192C}
{Cavuoti}, S., {Brescia}, M., \& {Longo}, G. 2014, ArXiv e-prints

\bibitem[{{Chen} \& {Chen}(2010)}]{2010ApJ...721.1790C}
{Chen}, C.~W. \& {Chen}, W.~P. 2010, \apj, 721, 1790

\bibitem[{{Colak} \& {Qahwaji}(2009)}]{2009SpWea...7.6001C}
{Colak}, T. \& {Qahwaji}, R. 2009, Space Weather, 7, 6001

\bibitem[{Cortes \& Vapnik(1995)}]{svm}
Cortes, C. \& Vapnik, V. 1995, Machine Learning, 20, 273

\bibitem[{{Daigle} {et~al.}(2003){Daigle}, {Joncas}, {Parizeau}, \&
  {Miville-Desch{\^e}nes}}]{2003PASP..115..662D}
{Daigle}, A., {Joncas}, G., {Parizeau}, M., \& {Miville-Desch{\^e}nes}, M.-A.
  2003, \pasp, 115, 662

\bibitem[{{Davoust}(1986)}]{1986A&A...166..177D}
{Davoust}, E. 1986, \aap, 166, 177

\bibitem[{{Djorgovski} {et~al.}(2012){Djorgovski}, {Mahabal}, {Donalek},
  {Graham}, {Drake}, {Moghaddam}, \& {Turmon}}]{2012arXiv1209.1681D}
{Djorgovski}, S.~G., {Mahabal}, A.~A., {Donalek}, C., {et~al.} 2012, ArXiv
  e-prints

\bibitem[{{du Buisson} {et~al.}(2014){du Buisson}, {Sivanandam}, {Bassett}, \&
  {Smith}}]{2014arXiv1407.4118D}
{du Buisson}, L., {Sivanandam}, N., {Bassett}, B.~A., \& {Smith}, M. 2014,
  ArXiv e-prints

\bibitem[{{Geach}(2012)}]{2012MNRAS.419.2633G}
{Geach}, J.~E. 2012, \mnras, 419, 2633

\bibitem[{{Gerdes} {et~al.}(2010){Gerdes}, {Sypniewski}, {McKay}, {Hao},
  {Weis}, {Wechsler}, \& {Busha}}]{2010ApJ...715..823G}
{Gerdes}, D.~W., {Sypniewski}, A.~J., {McKay}, T.~A., {et~al.} 2010, \apj, 715,
  823

\bibitem[{{Harris}(1996)}]{1996AJ....112.1487H}
{Harris}, W.~E. 1996, \aj, 112, 1487

\bibitem[{Hastie {et~al.}(2001)Hastie, Tibshirani, \&
  Friedman}]{hastie01statisticallearning}
Hastie, T., Tibshirani, R., \& Friedman, J. 2001, The Elements of Statistical
  Learning, Springer Series in Statistics (New York, NY, USA: Springer New York
  Inc.)

\bibitem[{{Heggie} \& {Mathieu}(1986)}]{1986LNP...267..233H}
{Heggie}, D.~C. \& {Mathieu}, R.~D. 1986, in Lecture Notes in Physics, Berlin
  Springer Verlag, Vol. 267, The Use of Supercomputers in Stellar Dynamics, ed.
  P.~{Hut} \& S.~L.~W. {McMillan}, 233

\bibitem[{{Hoyle} {et~al.}(2015{\natexlab{a}}){Hoyle}, {Rau}, {Paech},
  {Bonnett}, {Seitz}, \& {Weller}}]{2015arXiv150308214H}
{Hoyle}, B., {Rau}, M.~M., {Paech}, K., {et~al.} 2015{\natexlab{a}}, ArXiv
  e-prints

\bibitem[{{Hoyle} {et~al.}(2015{\natexlab{b}}){Hoyle}, {Rau}, {Zitlau},
  {Seitz}, \& {Weller}}]{2015MNRAS.449.1275H}
{Hoyle}, B., {Rau}, M.~M., {Zitlau}, R., {Seitz}, S., \& {Weller}, J.
  2015{\natexlab{b}}, \mnras, 449, 1275

\bibitem[{{King}(1966)}]{1966AJ.....71...64K}
{King}, I.~R. 1966, \aj, 71, 64

\bibitem[{{Kuminski} {et~al.}(2014){Kuminski}, {George}, {Wallin}, \&
  {Shamir}}]{2014PASP..126..959K}
{Kuminski}, E., {George}, J., {Wallin}, J., \& {Shamir}, L. 2014, \pasp, 126,
  959

\bibitem[{{Lee} {et~al.}(2013){Lee}, {Han}, {Joo}, {Jang}, {Na}, {Okamoto},
  {Arimoto}, {Lim}, {Kim}, \& {Yoon}}]{2013ApJ...778L..13L}
{Lee}, Y.-W., {Han}, S.-I., {Joo}, S.-J., {et~al.} 2013, \apjl, 778, L13

\bibitem[{{Li} {et~al.}(2006){Li}, {Zhang}, {Zhao}, \&
  {Yang}}]{2006astro.ph.12749L}
{Li}, L., {Zhang}, Y., {Zhao}, Y., \& {Yang}, D. 2006, ArXiv Astrophysics
  e-prints

\bibitem[{{Mahabal} {et~al.}(2008){Mahabal}, {Djorgovski}, {Turmon}, {Jewell},
  {Williams}, {Drake}, {Graham}, {Donalek}, {Glikman}, \& {Palomar-QUEST
  Team}}]{2008AN....329..288M}
{Mahabal}, A., {Djorgovski}, S.~G., {Turmon}, M., {et~al.} 2008, Astronomische
  Nachrichten, 329, 288

\bibitem[{{McLaughlin} \& {van der Marel}(2005)}]{2005ApJS..161..304M}
{McLaughlin}, D.~E. \& {van der Marel}, R.~P. 2005, \apjs, 161, 304

\bibitem[{{Mihos} \& {Hernquist}(1994)}]{1994ApJ...437L..47M}
{Mihos}, J.~C. \& {Hernquist}, L. 1994, \apjl, 437, L47

\bibitem[{{Miocchi} {et~al.}(2006){Miocchi}, {Capuzzo Dolcetta}, {Di Matteo},
  \& {Vicari}}]{2006ApJ...644..940M}
{Miocchi}, P., {Capuzzo Dolcetta}, R., {Di Matteo}, P., \& {Vicari}, A. 2006,
  \apj, 644, 940

\bibitem[{{Ntampaka} {et~al.}(2015){Ntampaka}, {Trac}, {Sutherland},
  {Battaglia}, {P{\'o}czos}, \& {Schneider}}]{2015ApJ...803...50N}
{Ntampaka}, M., {Trac}, H., {Sutherland}, D.~J., {et~al.} 2015, \apj, 803, 50

\bibitem[{Quinlan(1986)}]{Quinlan86inductionof}
Quinlan, J.~R. 1986, MACH. LEARN, 1, 81

\bibitem[{Quinlan(1993)}]{Quinlan:1993:CPM:152181}
Quinlan, J.~R. 1993, C4.5: Programs for Machine Learning (San Francisco, CA,
  USA: Morgan Kaufmann Publishers Inc.)

\bibitem[{{R Core Team}(2014)}]{R}
{R Core Team}. 2014, R: A Language and Environment for Statistical Computing, R
  Foundation for Statistical Computing, Vienna, Austria

\bibitem[{{Shamir} {et~al.}(2013){Shamir}, {Holincheck}, \&
  {Wallin}}]{2013A&C.....2...67S}
{Shamir}, L., {Holincheck}, A., \& {Wallin}, J. 2013, Astronomy and Computing,
  2, 67

\bibitem[{{Singal} {et~al.}(2011){Singal}, {Shmakova}, {Gerke}, {Griffith}, \&
  {Lotz}}]{2011PASP..123..615S}
{Singal}, J., {Shmakova}, M., {Gerke}, B., {Griffith}, R.~L., \& {Lotz}, J.
  2011, \pasp, 123, 615

\bibitem[{{Stepinski} {et~al.}(2009){Stepinski}, {Mendenhall}, \&
  {Bue}}]{2009Icar..203...77S}
{Stepinski}, T.~F., {Mendenhall}, M.~P., \& {Bue}, B.~D. 2009, \icarus, 203, 77

\bibitem[{{Sugimoto} \& {Makino}(1989)}]{1989PASJ...41.1117S}
{Sugimoto}, D. \& {Makino}, J. 1989, \pasj, 41, 1117

\bibitem[{{Tagliaferri} {et~al.}(2003){Tagliaferri}, {Longo}, {Andreon},
  {Capozziello}, {Donalek}, \& {Giordano}}]{2003LNCS.2859..226T}
{Tagliaferri}, R., {Longo}, G., {Andreon}, S., {et~al.} 2003, Lecture Notes in
  Computer Science, 2859, 226

\bibitem[{{Thilker} {et~al.}(1998){Thilker}, {Braun}, \&
  {Walterbos}}]{1998A&A...332..429T}
{Thilker}, D.~A., {Braun}, R., \& {Walterbos}, R.~M. 1998, \aap, 332, 429

\bibitem[{{Thurl} \& {Johnston}(2002)}]{2002ASPC..265..337T}
{Thurl}, C. \& {Johnston}, K.~V. 2002, in Astronomical Society of the Pacific
  Conference Series, Vol. 265, Omega Centauri, A Unique Window into
  Astrophysics, ed. F.~{van Leeuwen}, J.~D. {Hughes}, \& G.~{Piotto}, 337

\bibitem[{{Toomre}(1977)}]{1977egsp.conf..401T}
{Toomre}, A. 1977, in Evolution of Galaxies and Stellar Populations, ed. B.~M.
  {Tinsley} \& R.~B.~G. {Larson}, D.~Campbell, 401

\bibitem[{{Toomre} \& {Toomre}(1972)}]{1972ApJ...178..623T}
{Toomre}, A. \& {Toomre}, J. 1972, \apj, 178, 623

\bibitem[{{van den Bergh}(1996)}]{1996ApJ...471L..31V}
{van den Bergh}, S. 1996, \apjl, 471, L31

\bibitem[{{Vander Plas} {et~al.}(2014){Vander Plas}, {Connolly}, \&
  {Ivezic}}]{2014AAS...22325301V}
{Vander Plas}, J., {Connolly}, A.~J., \& {Ivezic}, Z. 2014, in American
  Astronomical Society Meeting Abstracts, Vol. 223, American Astronomical
  Society Meeting Abstracts 223, 253.01

\bibitem[{{VanderPlas} {et~al.}(2012){VanderPlas}, {Connolly}, {Ivezic}, \&
  {Gray}}]{2012cidu.conf...47V}
{VanderPlas}, J., {Connolly}, A.~J., {Ivezic}, Z., \& {Gray}, A. 2012, in
  Proceedings of Conference on Intelligent Data Understanding (CIDU), pp.
  47-54, 2012., 47--54

\bibitem[{{VanderPlas} {et~al.}(2014){VanderPlas}, {Fouesneau}, \&
  {Taylor}}]{2014ascl.soft07018V}
{VanderPlas}, J., {Fouesneau}, M., \& {Taylor}, J. 2014, {AstroML: Machine
  learning and data mining in astronomy}, Astrophysics Source Code Library

\bibitem[{{Varri} \& {Bertin}(2009)}]{2009ApJ...703.1911V}
{Varri}, A.~L. \& {Bertin}, G. 2009, \apj, 703, 1911

\bibitem[{{Varri} \& {Bertin}(2012)}]{2012A&A...540A..94V}
{Varri}, A.~L. \& {Bertin}, G. 2012, \aap, 540, A94

\bibitem[{{Vesperini} {et~al.}(2014){Vesperini}, {Varri}, {McMillan}, \&
  {Zepf}}]{2014MNRAS.443L..79V}
{Vesperini}, E., {Varri}, A.~L., {McMillan}, S.~L.~W., \& {Zepf}, S.~E. 2014,
  \mnras, 443, L79

\bibitem[{{White} \& {Shawl}(1987)}]{1987ApJ...317..246W}
{White}, R.~E. \& {Shawl}, S.~J. 1987, \apj, 317, 246

\bibitem[{{Wright} {et~al.}(2015){Wright}, {Smartt}, {Smith}, {Miller},
  {Kotak}, {Rest}, {Burgett}, {Chambers}, {Flewelling}, {Hodapp}, {Huber},
  {Jedicke}, {Kaiser}, {Metcalfe}, {Price}, {Tonry}, {Wainscoat}, \&
  {Waters}}]{2015MNRAS.449..451W}
{Wright}, D.~E., {Smartt}, S.~J., {Smith}, K.~W., {et~al.} 2015, \mnras, 449,
  451

\bibitem[{{Xu} {et~al.}(2013){Xu}, {Ho}, {Trac}, {Schneider}, {Poczos}, \&
  {Ntampaka}}]{2013ApJ...772..147X}
{Xu}, X., {Ho}, S., {Trac}, H., {et~al.} 2013, \apj, 772, 147

\bibitem[{{Yeche} {et~al.}(2009){Yeche}, {Petitjean}, {Rich}, {Aubourg},
  {Busca}, {Hamilton}, {Le Goff}, {Paris}, {Peirani}, {Pichon}, {Rollinde}, \&
  {Vargas-Magana}}]{2009arXiv0910.3770Y}
{Yeche}, C., {Petitjean}, P., {Rich}, J., {et~al.} 2009, ArXiv e-prints

\bibitem[{{Yu} {et~al.}(2009){Yu}, {Huang}, {Wang}, \&
  {Cui}}]{2009SoPh..255...91Y}
{Yu}, D., {Huang}, X., {Wang}, H., \& {Cui}, Y. 2009, \solphys, 255, 91

\end{thebibliography}
\bibliographystyle{aa}

\end{document}